\documentclass[12pt]{article}
\usepackage[T1]{fontenc}
\usepackage[latin2]{inputenc}
\usepackage{graphicx}

\title{Local density and the RPA corrections in charge
current quasielastic neutrino on Oxygen, Argon and Iron
scattering}

\author{Krzysztof M. Graczyk\footnote{\textit{e-mail
address: kgraczyk@ift.uni.wroc.pl}}\\ \\
Institute of Theoretical
Physics,
University of Wroc\l aw \\
pl. M. Borna 9,
50 - 204 Wroclaw\\
Poland }

\begin{document}

\maketitle


\abstract{ \footnotesize{Numerical computations of cross sections
for quasielastic charge current scattering of neutrino on Oxygen,
Argon and Iron in Local Density Approximation (LDA) are presented.
We consider processes for a few GeV neutrino energy. We include
corrections from nucleon re-interaction in nucleus described by
relativistic Random Phase Approximation (RPA). We adopt the
relativistic Fermi gas model of nucleus with and without taking
into account the effective mass of nucleons.} }

\section{Introduction}

The study of oscillations phenomena of neutrinos  became one of
the most interesting topics of particle physics. Experiments such
as K2K, MINOS, MiniBoone, ICARUS collect or will collect data
which allow to establish parameters of oscillations.  In the
analysis of experimental results  a good knowledge of theoretical
predictions of neutrino-nucleus cross sections is crucial. In long
baseline and atmospheric neutrino experiments one of the most
important  become analysis of the data for a few GeV
neutrinos\cite{Lipari}. In this energy domain the quasielastic
contribution is important.

The calculations of cross sections require a study of influence of
nuclear effects. The scattering amplitude can be combined from two
parts\cite{Walecka_old}. One is described by weak interaction
vertex of scattering neutrino on free nucleon and the other which
is evaluated in the framework of many body theory and describes
the model of nucleus. The weak interaction vertex is described
very accurately by Fermi theory  a few GeV neutrino. A
consideration of the model of nucleus described by Fermi gas (FG)
of nucleons\cite{Moniz} is one of the most popular and simple
starting point in discussions of nuclear effects in that energy
regime. In the Fermi gas model momenta of nucleons are uniformly
distributed inside the Fermi sphere whose radius is called Fermi
momentum - $k_F$. It is connected with baryon density inside
nucleus. For the first time this simple approach was applied in
the scattering of electrons on nuclei
\cite{Moniz-Whitney}-\cite{Donnelly}. Experimental results
obtained from  these processes allowed to establish average Fermi
momenta for given nuclear targets. Success of this approach caused
for application of it in neutrino-nucleus
interactions\cite{Smith_Moniz,Kuramoto,Engle,Horowitz_neutralne}.

As was mentioned above the Fermi gas model is the good starting
point for  further nuclear effects discussions. One can evaluate
this by consideration of some additional effects such as binding
energy, effective mass of nucleons,  final state interaction and
many others. However the most important seems to be taking into
account the collective behavior of nucleons inside the nucleus. It
can be done in several ways e.g. by using static potential, which
gets the quite easiest way to apply it or by application of the
methodology of quantum many body field theory.

One of the most consistent ways of describing of the  collective
properties of nucleons inside nucleus is the relativistic Mean
Field Theory (MFT)\cite{Walecka}. The ground state of nucleus is
described by relativistic Fermi gas of nucleons which has
effective mass introduced by self-consistency equation for given
target. The strong interaction between nucleons is introduced into
the model by consideration of virtual particles which are
exchanged between nucleons. They are described by the effective
interaction lagrangian which contain of vector and scalar fields.
The MFT allows to solve this theory. In the electron-nucleus
scattering, neutral $\sigma$ and vector $\omega$ are
considered\cite{Kurasawa}. The contribution from  $\sigma$ field
is fully manifested in the effective mass (the appearing of this
field lead to the  above mentioned self-consistency equation). In
the case of the charge current neutrino-nucleus interaction, pions
and mesons rho give contribution to scattering
amplitude\cite{Horowitz1,Mornas}. Corrections described by
effective lagrangian are calculated by using standard tools of
quantum perturbation field theory. It gives rise to infinite
number of diagrams.  It is impossible to take into account
contributions from all diagrams. One may chose special class of
them which correspond to  one particle - one hole excitations.
This approach is called relativistic Random Phase Approximation
(RPA) \cite{Feter}.

Experimental measurements of the electron-nucleus scattering show
that charge distribution in nucleus is not
uniform\cite{Sick_McCarthy,DeVries}, which means that baryon
density is not uniform either. The charge distribution can be
calculated in the framework of the relativistic MFT by solving the
Dirac-Hartree self-consistency equations\cite{Serot}. Solution of
it is consistent with experimental predictions. It allows us to
use  the experimental results and apply them to the formalism
under consideration.

The measurements of the scattering electron on nucleus allowed to
fit  charge density distributions inside nucleus. The fitting is
done by assumption that the simple models of nucleus such
oscillator model in the case of light nuclei e.g. Oxygen and Fermi
model in the case of heavier nuclei e.g. Argon, Iron. The
inclusion into the model of nucleus of the local profile of baryon
density (Local Density Approximation - LDA) is the treatment which
makes the model of nucleus more realistic. The effect of the local
density were taken into account not only in discussions of
electron nucleus data and in neutrino matter
interactions\cite{Singh}-\cite{Marteau}.

In this paper we want to develop consideration of previous
paper\cite{Graczyk}, where we adopted relativistic MFT. We
presented the analytical solution of the RPA Dayson equation and
discussed the RPA corrections in scattering of neutrino on
nucleus. Here we are going to extend our consideration by
application of Local Density Approach (LDA) in the cases of these
few targets. We will compare results for the case of LDA with
calculations which were done by assumption the constant baryon
density. The second case is described by average values of Fermi
momenta and effective masses for given targets. They were
calculated from experimental charge density
distributions\cite{DeVries}.

We choose  nuclei $_8O^{16}$, $_{18}{Ar}^{40}$ and
$_{26}{Fe}^{56}$, which characterize targets of neutrino
experiments: Super Kamionkande, Icarus and Minos. As mentioned
above, we are going to compare differential and total cross
sections for the nuclei with constant and local density. The
comparison be done in various levels of the model (Fermi gas with
or without RPA and with or without effective mass).

The paper is organized as follows. A short description of the
formalism is presented in section 2. In section 3 numerical
computations of cross sections are presented and the results are
discussed. In this section we also summarized.

\section{The formalism}

We consider the process:
$$
\nu_\mu + N(A,Z) \rightarrow \mu^- + p + N(A-1,Z)
$$
The cross section per one nucleon in the laboratory frame for the
above scattering is the following:
\begin{equation}
\frac{d^2 \sigma }{d|\vec{q}| \, dq_0}= -\frac{ G_F^2
\mathrm{cos}^2\theta_c \; |\vec{q}|}{16 \pi^2 \rho_F E^2 }
\mathrm{Im} \left ( {L_\mu}^\nu {\Pi^\mu}_\nu \right ).
\end{equation}
where $\rho_F =k_F^3/3\pi^2$, $q = k-k'$ is the transfer of
four-momentum, $k$, $k'$  are neutrino and lepton four momenta.
$L^{\mu\nu}$ is a leptonic tensor:
$$
L_{\mu\nu} = 8 \left ( k_\mu {k'}_\nu + {k'}_\mu k_\nu -
g_{\mu\nu}k_\alpha{k'}^\alpha \pm
\mathrm{i}\epsilon_{\mu\nu\alpha\beta}{k'}^\alpha k^\beta \right
).
$$
The $\pm$ sign  depends on the considered process
(neutrino/antineutrino). Nuclear properties of nucleus are
included in the polarization tensor $\Pi^{\mu\nu}$. This tensor is
evaluated by the QFT techniques \cite{Feter} and can be written as
the sum of two contributions. One of them is  the free Fermi gas
and second the RPA correction:
$$
\Pi^{\mu\nu}(q)= \Pi^{\mu\nu}_{free}(q) +
\Pi^{\mu\nu}_{\mathrm{RPA}}(q),
$$
\begin{equation}
\label{free_Pi} \Pi_{free}^{\mu\nu}(q)  = -\mathrm{i} \int
\frac{d^4 p}{( 2 \pi )^4} \mathrm{Tr} \left ( \Delta_{FG}(p +
q)\Gamma^\mu(q)\Delta_{FG}(p)\Gamma^\nu (-q ) \right ).
\end{equation}
By  $\Gamma^\mu$  elementary weak charge current nucleon-nucleon
is described and expressed by means of the form factors \cite{LS}:
\begin{equation}
\Gamma^\alpha(q_\mu) = F_1(q^2_\mu)\gamma^\alpha
+F_2(q_\mu^2)\frac{\mathrm{i}\sigma^{\alpha\nu}q_\nu}{2M} +
G_A(q_\mu^2)\gamma^\alpha\gamma^5
\end{equation}
We omit $G_p(q_\mu)$ term since its contribution to $\nu_{e}$ and
$\nu_{\mu}$ cross sections at $E_{\nu}\sim 1GeV$ is negligible.
The value of parameters which characterize the form factors are
presented in the appendix.

By the $\Delta_{FG}(p)$  in (\ref{free_Pi}) we denote the
propagators of nucleon in  the free Fermi sea:
\begin{equation}
 \Delta_{FG}(p) = (p \!\!\! / + M^* ) \left ( \frac{ 1 }{
p_\alpha^2 - {M^*}^2 + \mathrm{i}\epsilon } +
\frac{\mathrm{i}\pi}{E_p}\delta(p_0 - E_p)\theta(k_F - p) \right
).
\end{equation}
$M^*$ is  the effective mass, calculated from the self-consistency
equation for given Fermi momentum $k_F$ \cite{Walecka}:
\begin{eqnarray}
\label{self-cosistency} M^* & = &
M-\frac{g_s^2}{m_s^2}\frac{M^*}{\pi^2} \left( k_F E_F^* - M^*
\mathrm{ln}\left(\frac{k_F+E_F^*}{M^*}\right)\right),\\
E_F & =&\left(k_F^2+M^{*2}\right)^{1/2} .\nonumber
\end{eqnarray}
In our calculations we will separate the case with the free
nucleon mass $M^*=M=939$ MeV and with the effective mass.

The RPA part of the polarization tensor is calculated by
consideration of contributions from one particle - one hole
excitations. It leads to the correspond Dyson equation whose
solution gives the $\Pi_{RPA}$(for more details see
\cite{Graczyk}).

\subsection{Local Density Approximation}

We are going to compare cross sections for the case of the model
of nucleus with constant baryon density with the case  of the
model of nucleus with local baryon density. The constant value of
baryon density for given nucleus is evaluated by calculation of
average value of Fermi momentum for given density distribution
\begin{eqnarray}
\label{sredni_kf} <k_F> & = & \frac{\int d^3 r
k_F(r)\rho(r)}{A}, \\
{k_F}_{p}(r) & = & \sqrt[3]{3\pi^2\frac{Z}{A}\rho(r)},\\
{k_F}_{n}(r) & = &\sqrt[3]{3\pi^2\frac{N}{A}\rho(r)}.
\end{eqnarray}
$A=\int d^3 r \rho(r)$  is a atomic number, $\rho(r)$ - charge
density profile\cite{DeVries}.

The results for  the local baryon density are obtained by the
integration of the cross section per one nucleon with weight given
by density.
\begin{equation}
d\sigma_{local\; density} = 4\pi \int_0^{\mathrm{cutoff}} dr r^2
\rho_{n,p}(r)
d\sigma(k_F(r),M^*(r))|_{\mathrm{per}\,\mathrm{nucleon}},
\end{equation}
where the index of n or p in $\rho$ corresponds to neutron or
proton density distribution.

\begin{figure}[ht]
\centerline{
\includegraphics[width=10cm]{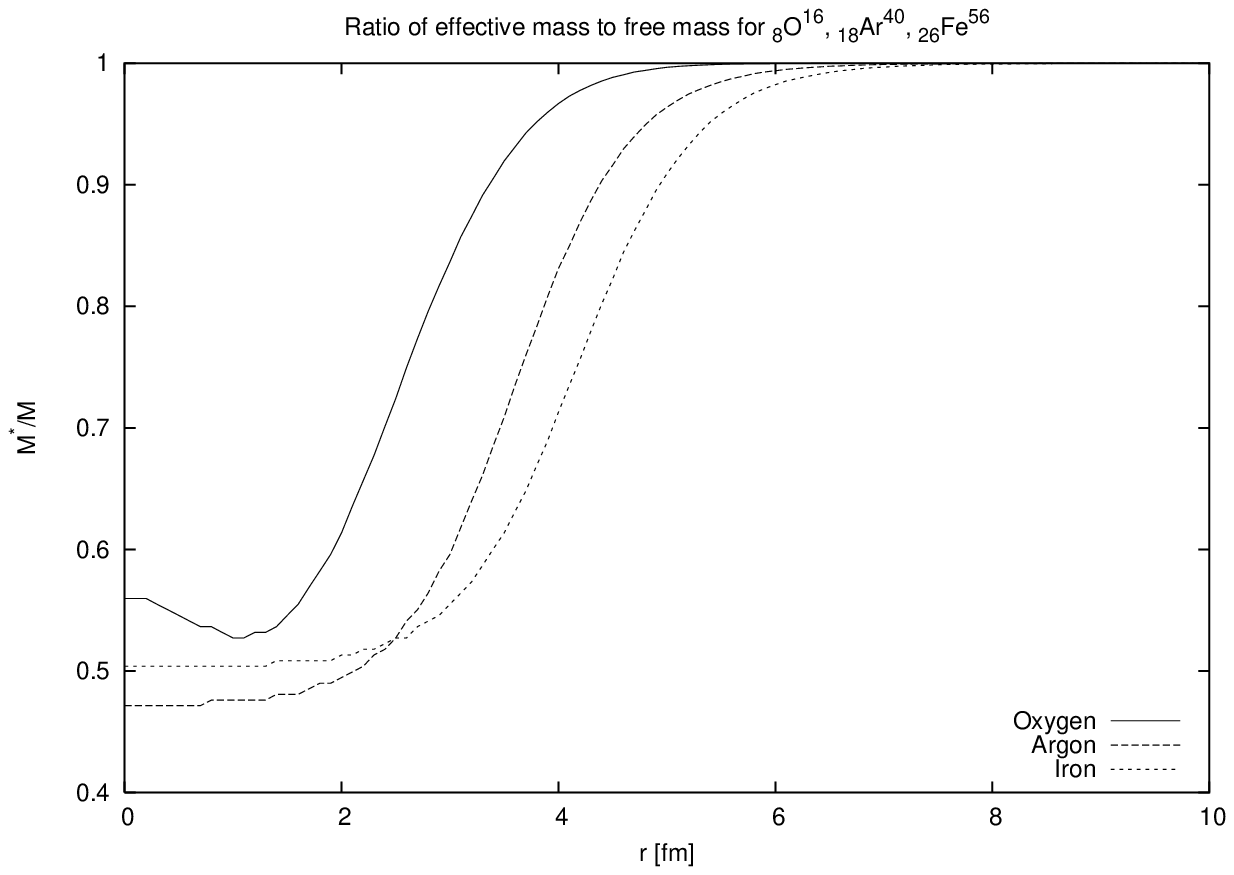}
} \caption{\scriptsize The ratio of effective mass to the free
nucleon mass calculated by solving of self-consistency  equation
(\ref{self-cosistency}) for a local value $k_F(r)$. }
\label{wykres0}
\end{figure}

\begin{table}[ht]
  \centering
\begin{tabular}{||l|c|c||}
 \hline\hline
  Nucleus &  $<k_F>$ [MeV] & $<M^*> [MeV]$
  \\ \hline
  Oxygen & 199.21 & 690.78 \\ \hline
  Argon  & 216.88 & 631.37 \\ \hline
  Iron  & 216.91 & 634.84 \\ \hline
\end{tabular}
\caption{\scriptsize The average values of Fermi momenta and
corresponding effective masses calculated from charge
distributions for nuclei.  }\label{tabela1}
\end{table}
In presented approach, for each values of integrating variable $r$
the Fermi momenta $k_F(r)$ and effective masses $M^*(r)$ are
calculated. It is done numerically. The reconstructed
distributions of the effective mass inside nuclei are presented in
fig. \ref{wykres0}. We also calculate the average Fermi momenta
for Oxygen, Argon and Iron as well as effective masses. The
results are presented in table \ref{tabela1}.

\section{Numerical results and discussion}

We show a set of plots which present the  differential cross
sections for 1 GeV neutrino energy. We calculated also the ratio
of total cross for the model of nucleus with constant density to
cross sections calculate in LDA at 1 GeV neutrino energy
(table.\ref{tabela2}). In the end we computed the total cross
sections for the three investigated targets. It was made by taking
into account all considered nuclear effects (LDA, RPA corrections
and the effective mass).
\begin{table}[ht]
  \centering
\begin{tabular}{||l|c|c|c|c||}
 \hline\hline
  Nucleus & FG & FG $(M^*)$ & FG + RPA & FG + RPA $(M^*)$
  \\ \hline
  Oxygen   & 0.997 & 0.994 & 0,997 & 0.993 \\ \hline
  Argon  & 0.994 & 0.986 & 0,994 & 0.989 \\ \hline
  Iron   & 0.994 & 0.987 & 0,995 & 0.990 \\ \hline
\end{tabular}
\caption{\scriptsize The ratio of  total cross sections calculated
for nucleus with local density to cross section for the nucleus
with constant density. The calculations were done for 1 GeV
neutrino energy. FG means the free Fermi gas model. The appearing
of $M^*$ means of application of the effective
mass.}\label{tabela2}

\end{table}

In the table \ref{tabela2} we made separation of several cases. We
computed ratio of cross sections for the free Fermi gas, the Fermi
gas with effective mass, the Fermi gas with RPA without and with
the effective mass. In the case of the model of nucleus with
constant density the average Fermi momenta and the effective
masses from table \ref{tabela1} were applied. As can be seen the
differences between approaches are in order of $0 \div 1.5 \% $.
It does not depend on how sophisticated complications the model of
nucleus is. One can also notice that the differences are slightly
bigger for argon and iron. These two targets are heavier and
charge distributions for them were extracted in the different way.

\begin{figure}[ht]
\centerline{
\includegraphics[width=10cm]{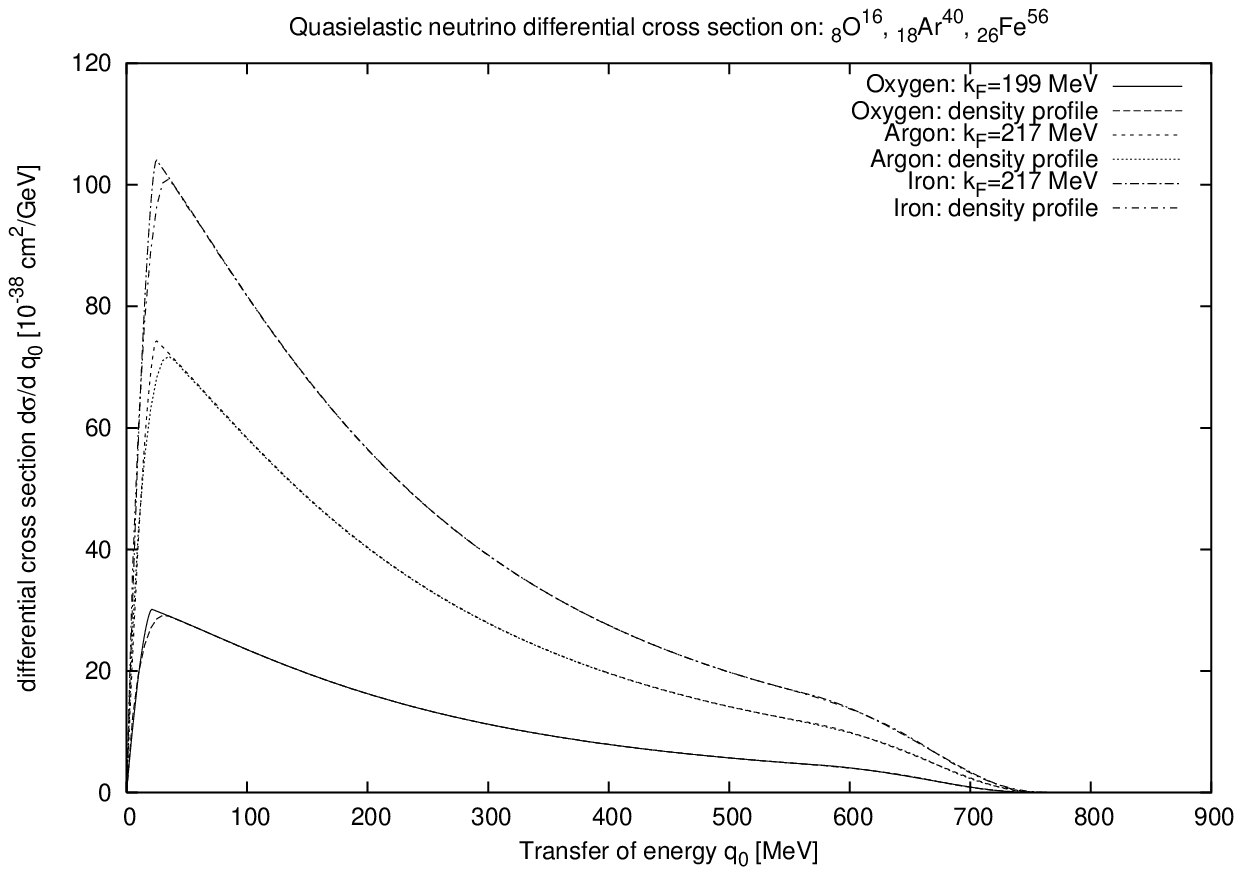}
} \caption{\scriptsize Comparison of differential cross sections
for the scattering of neutrinos  on nucleus with  local and
constant baryon density. The model of nucleus is   the free Fermi
gas ($M^* =939 MeV$). Computations were done for neutrino energy 1
GeV.}\label{wykres1}
\end{figure}

\begin{figure}[ht]
\centerline{
\includegraphics[width=10cm]{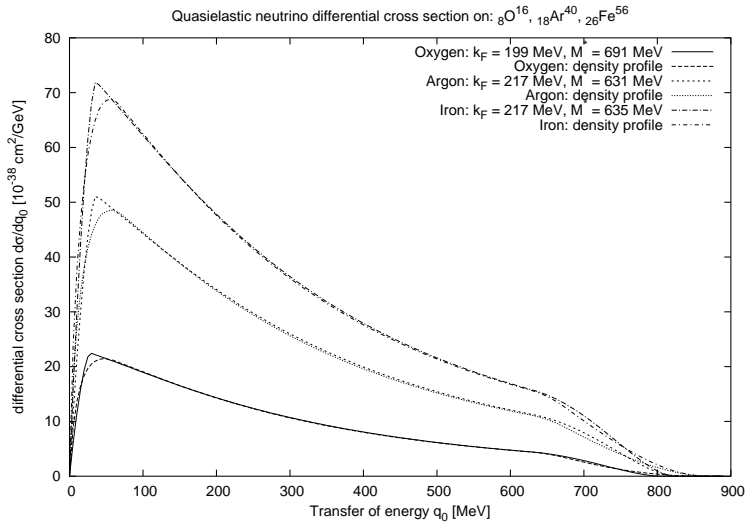}
} \caption{\scriptsize Comparison of differential cross sections
for the scattering of neutrinos  with nucleus with  local and
constant baryon density. The model of nucleus is the free Fermi
gas with effective mass. Computations were done for neutrino
energy  1 GeV. \label{wykres2} }
\end{figure}

Make our discussion more precise we present details comparison of
differential cross sections. In the figure \ref{wykres1} and
\ref{wykres2} the differential cross sections for scattering of 1
GeV neutrino on Oxygen, Argon and Iron are presented. It was
calculated for the Fermi gas model of nucleus with and without
adoption of the effective mass. The differences between approaches
with constant and local baryon density of nucleus are minor. One
can notice that only the maximum of the curve in the LDA is little
cut. The application of effective mass lowers the pick of cross
sections by about 30 $\%$ but also does not change the
differentiae between results for LDA and constant baryon density.

In last three figures we consider  the Fermi gas with corrections
from random phase approximation. RPA corrections are calculated by
counting infinite number of one loop digrams\cite{Horowitz1}. We
consider charge current reactions with mesons $\rho$ and pions
contributions into RPA polarization tensor. We applied standard
values of coupling constans for these fields. The short range
correlation between nucleons is applied by using Landau-Migdal
parameter $g'$ \cite{Horowitz1}. In our calculations we adopt
$g'=0.7$.

\begin{figure}[ht]
\centerline{
\includegraphics[width=10cm]{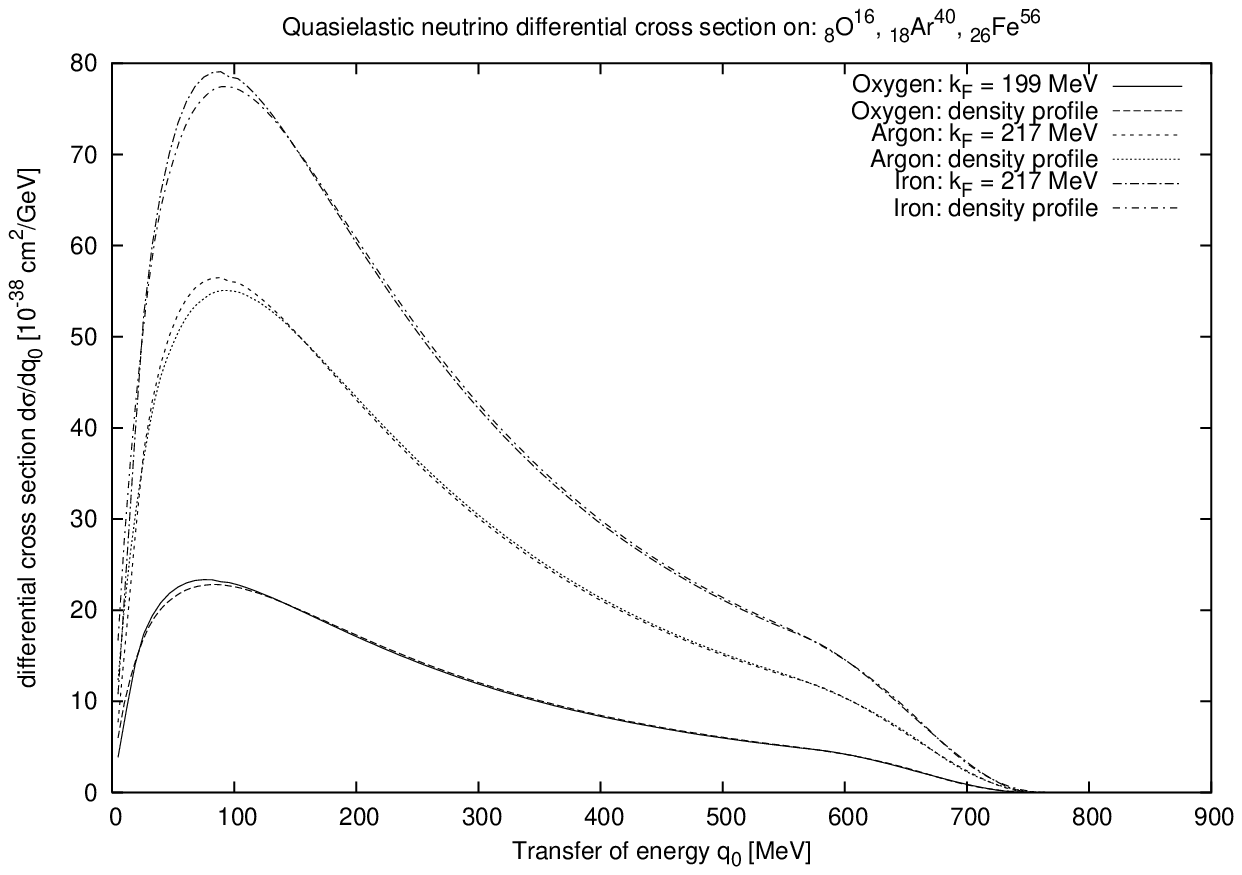}
} \caption{\scriptsize Comparison of differential cross sections
for the scattering of neutrinos  with nucleus with  local and
constant baryon density. The model of nucleus is  by  Fermi gas
with RPA corrections ($M^* =939 MeV$). Computations were done for
neutrino energy 1 GeV. \label{wykres3} }
\end{figure}

\begin{figure}[ht]
\centerline{
\includegraphics[width=10cm]{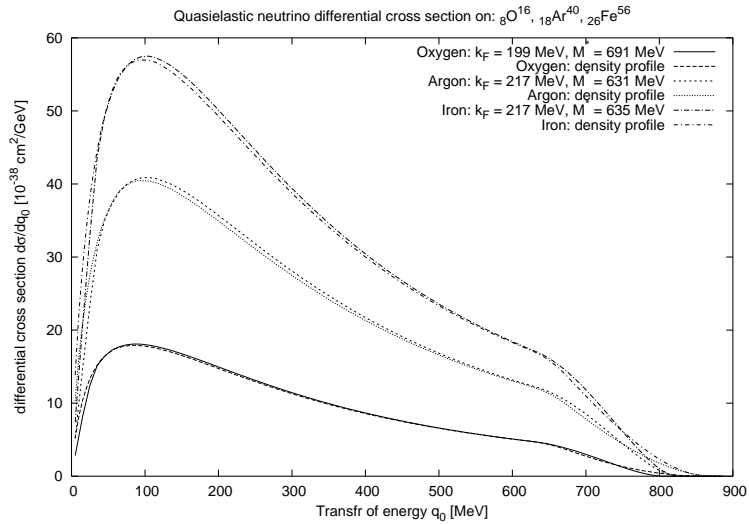}
} \caption{\scriptsize Comparison of differential cross sections
for the scattering of neutrinos  with nucleus with  local and
constant baryon density. The model of nucleus is  the Fermi gas
with RPA corrections and  effective mass. Computations were done
for neutrino energy  1 GeV.\label{wykres4} }
\end{figure}

\begin{figure}[ht]
\centerline{
\includegraphics[width=10cm]{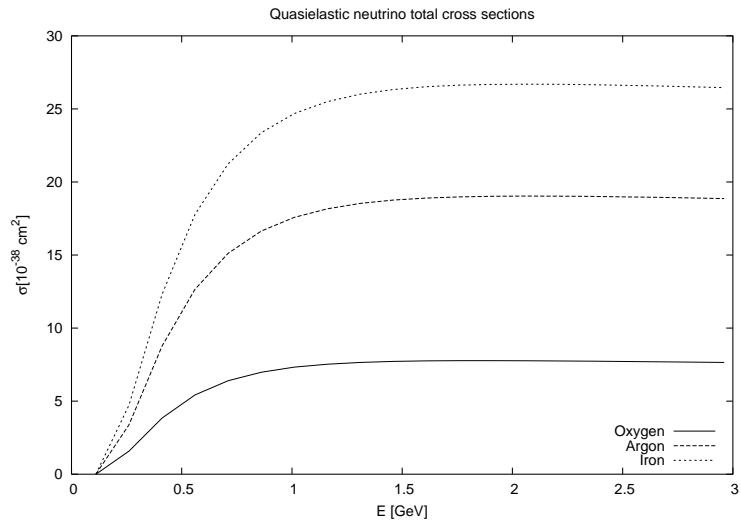}
} \caption{\scriptsize Total  cross sections for scattering of
neutrino on Oxygen, Argon and Iron. Calculations were done by
applying corresponding charge densities profiles($k_F$ and $M^*$
are local) and RPA. \label{wykres5} }
\end{figure}

In figure \ref{wykres3} we  present calculations with the RPA
corrections and  $M^* = M = 939 $ MeV. Similarly, as was noticed
in the previous case (fig.\ref{wykres1}), only small differences
in the maximum are observed.  The inclusion of the effective mass
in the RPA calculations dose not change the LDA effect (fig.
\ref{wykres4}). When we compare the figure \ref{wykres2} with fig.
\ref{wykres3} and figure \ref{wykres4} with figre \ref{wykres5}
one can see that the RPA effect decrease the cross sections,
especially for small transfers of the energy.

We conclude that local density effect could be replaced by
calculation of equivalent average parameters of the model which
make the numerical calculations faster. The application of the
approach does not depend on complication introduced dynamical
model of reaction.

\appendix

\section{Appendix}
\begin{itemize}
\item\textbf{Form Factors}
\end{itemize}
In our computations we used dipol form factors\cite{LS} with: $M_A
= 1.03 \;\mathrm{GeV}$ , $ M_V^2 = 0.71 \,\mathrm{GeV}^2$, $\mu =
4.71 $ and $g_A = G_A(0) =  -1.26$.

\begin{itemize}
\item\textbf{Charge density distributions}
\end{itemize}
We adopt the charge density  profiles from ref. \cite{DeVries}.

Charge density profile for Oxygen $_8 O^{16}$ is determined by
harmonic oscillator model
\begin{equation}
\label{density_O}
\rho(r)=\rho(0)\exp\left(-\frac{r^2}{R^2}\right)
\left(1+C\left(\frac{r}{R}\right)^2+C_1\left(\frac{r}{R}\right)^4\right)
\end{equation}
where:
$$\rho(0)=0.141 \,\rm{fm}^{-3}\;\;\; R=1.833\,\rm{fm}\;\;\;C=1.544.\;\;\;C_1=0.$$
In the case of $_{18} Ar^{40}$ and $_{26} Fe^{56}$ the two
parameters Fermi model gives the following profiles:
\begin{equation}
\label{density_ArFe} \rho(r)=\frac{\rho(0)}{\left(
1+\exp\left((r-c)/C_1\right)\right)}.
\end{equation}
in the case of Argon:
$$\rho(0)=0.176\,\rm{fm}^{-3},\;\;\; C=3.530\,\rm{fm},\;\;\;
C_1=0.542\,\rm{fm}.$$ in the case of Iron:
$$\rho(0)=0.163\,\rm{fm}^{-3},\;\;\; C=4.111\,\rm{fm}, \;\;\;C_1=0.558\,\rm{fm}.$$

\textbf{\large Acknowledgments}

This work was supported by KBN grant 105/E-344/SPB/ICARUS/P-03/DZ
211/2003-2005.  I would like to thank Prof. Jan Sobczyk for useful
discussions, remarks and suggestions which resulted in improvement
of this paper. I thank also to C. Juszczak and J. Nowak for
several comments.

\newpage

\end{document}